\documentclass[preprint]{aastex}
\usepackage{mkfig}

\begin{document}

\title{\bf Interpreting the Mg~II h and k Line Profiles of Mira Variables}

\author{B. E. Wood\altaffilmark{1}, M. Karovska}
\affil{Harvard-Smithsonian Center for Astrophysics, 60 Garden St., Cambridge,
  MA 02138.}
\email{wood@head-cfa.harvard.edu, karovska@head-cfa.harvard.edu}

\altaffiltext{1}
  {Present address: JILA, University of Colorado, Boulder, CO 80309-0440.}

\begin{abstract}

     We use radiative transfer calculations to reproduce the basic
appearance of Mg~II lines observed from Mira variables.  These lines have
centroids that are blueshifted by at least 30 km~s$^{-1}$ from the stellar
rest frame.  It is unlikely that flow velocities in the stellar atmospheres
are this fast, so radiative transfer effects must be responsible for this
behavior.  Published hydrodynamic models predict the existence of
cool, downflowing material above the shocked material responsible for the
Mg~II emission, and we demonstrate that scattering in this layer can result
in Mg~II profiles as highly blueshifted as those that are observed.  However,
our models also show that scattering {\em within} the shock plays an
equally strong role in shaping the Mg~II profiles, and our calculations
illustrate the importance of partial redistribution and the effects of
being out of ionization equilibrium.

\end{abstract}


\section{Introduction}

     Mira variables are an important class of pulsating variable stars
representing the final stages in the life of a solar-type star.  The stellar
pulsation of these stars drives periodic shock waves
through their atmospheres.  These shocks waves determine the
atmospheric structure of Miras to a large extent, and they assist in
producing very high mass loss rates.  The Mg~II h \& k lines of Mira
variables at 2802.705~\AA\ and 2795.528~\AA, respectively, are useful
diagnostics for the shocks, since they are produced within the heated
plasma just behind the outwardly-propagating shocks.

     \citet[][hereafter Paper 1]{bew00} analyzed IUE observations of
several extensively observed Mira variables in order to study the properties
of their UV emission lines, especially Mg~II h \& k.  Figure 1 shows a
sequence of IUE spectra of the Mira variable R~Car, illustrating how the
Mg~II h line profile typically varies during the course of a pulsation cycle.
The Mg~II flux rises well after optical maximum ($\phi=0.0$), peaking near
$\phi=0.2-0.5$ and then decreasing \citep[see also][]{ewb90,dgl96}.
Although this pattern is observed for every pulsation
cycle, the amount of Mg~II flux produced during a pulsation
cycle can vary greatly from one cycle to the next.

     The Mg~II k lines are almost always contaminated with absorption lines
of Fe~I and Mn~I, indicating the existence of overlying cool material above
the Mg~II emission region.  The Mg~II h line, however, is not greatly
contaminated by such absorption.  Thus, most of this paper (and Paper 1)
focuses on the more pristine h line profile.

     The centroid of the Mg~II h line is always highly blueshifted by at
least 30 km~s$^{-1}$ from the rest frame of the star, and the magnitude of
the blueshift is clearly observed to decrease with pulsation phase, as seen
in Figure 1.  The blueshifts vary somewhat from star to star and cycle to
cycle, but typical centroid changes are from $-70$ km~s$^{-1}$ to
$-40$ km~s$^{-1}$ from $\phi=0.2$ to $\phi=0.6$ (see Paper 1).  This
behavior is very similar to that of the optical Ca~II H \& K lines
\citep{pwm60}.  The width of the Mg~II h line is also phase-dependent,
decreasing from about 70 km~s$^{-1}$ to about 40 km~s$^{-1}$ between
$\phi=0.2$ and $\phi=0.6$ (see Fig.\ 1).

     In addition to Mg~II h \& k, many multiplets of Fe~II lines are
observed in the UV spectra of Miras.  Another useful line detected in
these data is the Al~II] $\lambda$2669 line.  The fluxes of the Fe~II and
Al~II] lines exhibit the same phase-dependent behavior as the Mg~II lines,
but their widths and centroid velocities are of a different character.
The Fe~II and Al~II] lines have very narrow widths which are not
phase-dependent.  Except for Fe~II $\lambda$2599 (see below), the Fe~II and
Al~II] lines of most of the Miras show modest blueshifts of
$5-15$ km~s$^{-1}$, which are not variable within the error bars.  In this
paper, we use simple radiative transfer calculations of Mg~II line profiles
to try to explain why the Mg~II lines are broader and have much larger
blueshifts than the less optically thick Fe~II and Al~II] lines.

\section{Possible Explanations for the Mg~II Line Shifts}

     The Fe~II and Al~II] lines will have significantly lower opacities
than the much stronger Mg~II h \& k lines, and some may even be optically
thin \citep{kgc95}.  This opacity difference is presumably
the primary reason for the behavioral differences of these lines.
Evidence for this is provided by the most optically thick Fe~II line at
2599.394~\AA, which behaves differently from the other Fe~II lines and in
fact exhibits some of the characteristics of Mg~II h \& k, with larger line
widths and large blueshifts (Paper 1).

     Because of their low opacity, the Fe~II and Al~II] line centroids
should be more indicative of the true outflow velocities of the shocked
material, and the line widths should be more indicative of turbulent
velocities within the shocked material.  In the following, we explore
possible reasons why the Mg~II lines are broader and more blueshifted.

     Models and observations of various stellar sources demonstrate that as
the opacity in an emission line is increased, the line typically broadens and
separates into two peaks \citep[e.g.][]{rdr98}.  This is why solar
Lyman-$\alpha$, Mg~II h \& k, and Ca~II H \& K lines are observed to be
double-peaked \citep{uf77}.  Many Mg~II lines of
red giant and supergiant stars are also double-peaked, although the blue
peak is often affected by wind absorption \citep{rdr95}.
For Miras, perhaps we are seeing only the blue peak of the Mg~II emission,
with the red peak being suppressed by overlying material.

     One possible reason why only the blue peak is visible is
that circumstellar neutral material is completely
absorbing the red peak.  The observed Mg~II k lines
of Miras are clearly affected by this overlying neutral material
(see Paper 1).  However, this interpretation requires far more neutral
absorption on the red side of the line
than on the blue side for both the h and k lines.  This would also have
to be the case for the Ca~II H \& K lines, which behave similarly to Mg~II
h \& k; and also for the Fe~II $\lambda$2599 line to
explain its larger blueshift.  It seems very unlikely that neutral absorption
features would blanket predominantly the red sides of {\em all} these
lines.  The suppression of the red peak must therefore be due to radiative
transfer processes within the lines themselves.

     The simplest explanation is that
there is downflowing Mg~II material suppressing the emission produced by
outflowing Mg~II material.  Hydrodynamic models of pulsating Miras suggest
that after material is accelerated outwards by shocks it will gravitationally
decelerate and fall back towards the star, meaning there should be
downflowing material present above the pulsation-induced shocks
\citep{ghb88,msb96,sh98}.

     \citet{rpk57} used such a
scenario to try to explain the behavior of Ca~II H \& K lines observed from
Cepheids and Miras.  He assumed that an overlying absorption layer
of downflowing Ca~II material obliterates the red side of the Ca~II
lines.  One problem with this simple model is that unless densities in the
downflowing material are many orders of magnitude higher than one would
expect based on the hydrodynamic models, the layer should be a scattering
layer and not an absorption layer.  At the relatively low densities that
are expected to exist in this region of the atmosphere, there is no
immediately apparent way to destroy/absorb Mg~II photons.  In principle, a
highly opaque scattering layer might also work, but in a scattering
environment it is harder to completely suppress the entire red side of a
line since photons can frequency scatter far enough into the red wing of the
line to escape the layer.  In order to test whether a reasonable
downflowing scattering layer can in fact suppress the red peaks of the Mg~II
lines, we perform the following radiative transfer simulation.

\section{Radiative Transfer Calculations}

\subsection{Methodology}

     The hydrodynamic models of pulsating Miras show that the pulsation
produces a series of outwardly-propagating shocks within the outer
atmosphere of these stars.  In Figure 2a-c, we show schematic density,
temperature, and velocity profiles that represent the innermost shock of a
pulsating Mira and the region overlying it, which extends nearly to the next
shock.  These curves are meant to be crudely similar to the model results of
\citet{ghb88} and \citet{law88} at a pulsation phase of about $\phi=0.5$.
In practice, we experimented with minor adjustments to all these curves in
order to find a model that adequately reproduced the observations.  The
density, temperature, and velocity structure of the atmosphere all play a
role in determining the properties of the Mg~II profiles.

     We computed the ionization state of the model atmosphere in Figure 2
using collisional ionization rates from \citet{gsv97} and recombination
rates from \citet{jms82}.  In computing the electron density
we consider all abundant elements whose ionization might contribute
significant numbers of electrons.  We assume the solar abundances of
\citet{ea89} for these calculations.  Magnesium, hydrogen,
silicon, and iron turn out to be the dominant electron contributors.

     The thin lines in Figure 2d show the electron density and Mg~II density
computed assuming ionization equilibrium.  However, ionization equilibrium is
not a good approximation for this dynamic atmosphere.  Typical atomic
recombination rates are $\alpha \sim 10^{-12}$ cm$^{3}$ s$^{-1}$ for the
temperatures within the model atmosphere.  The recombination timescale is
$t_{r}=(n_{e}\alpha)^{-1}$.  For electron densities below
$n_{e}=3\times 10^{4}$ cm$^{-3}$, $t_{r}$ is longer than most Mira pulsation
periods.  Recombination will not have time to
occur when the electron density approaches this value, so the plasma outside
the shocks will be frozen in a much higher ionization state
than suggested by the ionization equilibrium calculations (see Fig.\ 2d).

     In order to estimate more reasonable electron and Mg~II densities, we
performed the following simulation.  Ignoring the velocity structure
implied by Figure 2c, we followed the ionization state of a static parcel
of gas as we repeatedly ran the density and temperature profiles of Figures
2a-b over the parcel with a period of 300 days, a typical Mira pulsation
period.  The electron density of the parcel was initially set to some
arbitrary value.  At the temperatures with which we are concerned, we can
assume each atom is either neutral or singly-ionized.  Collisional
ionization and radiative recombination are the two processes responsible for
establishing the ionization state of each element, and the equation
\begin{equation}
\frac{dn_{2}}{dt}=n_{e}\left( n_{1}C_{12} - n_{2}\alpha_{21} \right)
\end{equation}
shows how the density of the ionized state of the element in question
($n_{2}$) changes due to these processes, where $n_{1}$ is the density of
the neutral state of the element, and $C_{12}$ and $\alpha_{21}$ are the
temperature dependent ionization and recombination rates to and from the
singly ionized state, respectively.  By solving this simple differential
equation, we can express the time dependent density of the singly-ionized
state of a given element after a time step $\Delta t$ as
\begin{equation}
n_{2}(t+\Delta t)=\frac{n_{tot}C_{12}}{C_{12}+\alpha_{21}} + \left( n_{2}(t) -
  \frac{n_{tot}C_{12}}{C_{12}+\alpha_{21}} \right) \exp \left( -n_{e}(t)
  \left( C_{12}+\alpha_{21} \right) \Delta t \right),
\end{equation}
where $n_{tot}$ is the total number density of the element
($n_{tot}=n_{1}+n_{2}$).

     The ionization states of each element, and the electron density, are
recomputed after each time step.  After just a few pulsation cycles, the
electron density and ionization states settle into a repeatable pattern,
from which we can extract the electron and Mg~II density profiles shown as
thick lines in Figure 2d.  As expected, these densities are very different
from the equilibrium values.

     We now have all the information we need to compute theoretical Mg~II
h \& k line profiles.  Since the observed k line is usually contaminated with
neutral absorption features (see Paper 1), we focus on the h line.  We
use a simple two-level atom approximation for our calculations.  At each
level of the atmosphere the number of h line photons generated by collisional
excitation per unit volume per unit time is then given (in cgs units) by
\citep[e.g.,][]{ab84}
\begin{equation}
L=\frac{8.6\times 10^{-6} \Omega n_{e} n}{g T^{0.5}}
  \exp \left( -\frac{1.44\times 10^{8}}{\lambda T} \right),
\end{equation}
where $n$ is the Mg~II number density, $\Omega$ is the collision
strength, $g$ is the statistical weight of the lower level, and $\lambda$ is
the wavelength of the transition in {\AA}ngstroms.  For the h line,
$\Omega=5.6$ and $g=2$ \citep{cm81}.

     The wavelength dependent optical depth for each level of the
atmosphere is \citep[e.g.,][]{ls78}
\begin{equation}
\tau_{\lambda}(h)=0.02654 f \int_{h}^{h_{top}} n(h) \phi_{\lambda}(h) dh,
\end{equation}
where $f$ is the oscillator absorption strength, $n(h)$ is the height
dependent Mg~II density, $h_{top}$ is the height at the top of the atmosphere,
and $\phi_{\lambda}$ is the line profile function.  For the Mg~II h line,
$f=0.3054$ \citep{dcm91}.  The assumed line profile is a Voigt profile with a
centroid defined by the velocity profile in Figure 2c, and a width defined by
a height-dependent microturbulence function, $b(h)$, also shown in Figure 2c.
As the figure shows, we assume a microturbulence of 5 km~s$^{-1}$ inside the
shock and 3 km~s$^{-1}$ outside, although in practice we experiment with
different values (see below).

     Because the critical density where the collisional deexcitation rate
equals that of radiative deexcitation ($n_{crit}\approx 5\times 10^{14}$
cm$^{-3}$) is at least $10^{9}$ times higher than the densities in our model
atmosphere, we neglect collisional deexcitation in our calculations, making
this truly a pure scattering atmosphere.  We assume complete angular
redistribution for each scattering event, but for the frequency
redistribution we experiment with both complete redistribution (CRD) and
partial redistribution (PRD).  In the PRD treatment, the redistribution
function defining the probability of a photon absorbed with frequency $\nu$
being reemitted with frequency $\nu^{\prime}$ in a comoving rest frame can
be expressed as \citep[e.g.][]{rwm73,gsb80}
\begin{equation}
R(\nu,\nu^{\prime})=(1-\Lambda)R_{II}(\nu,\nu^{\prime})+
  \Lambda R_{III}(\nu,\nu^{\prime}),
\end{equation}
where $R_{II}$ and $R_{III}$ are the integral functions first defined by
\citet{dgh62}.  The $\Lambda$ paramenter is
\begin{equation}
\Lambda = \frac{\Gamma_{col}}{\Gamma_{col}+\Gamma_{rad}},
\end{equation}
where $\Gamma_{col}$ and $\Gamma_{rad}$ are the collisional and radiative
damping parameters, respectively.  However, for our model atmosphere
collision rates are very low because densities are very low.  As a
consequence, $\Gamma_{col}$ and $\Lambda$ are both very small; we estimate
$\Lambda \sim 10^{-8}$ at most.  Thus, for our purposes, 
$R(\nu,\nu^{\prime})\approx R_{II}(\nu,\nu^{\prime})$ is a very good
approximation.

     Several techniques have been used in the past to evaluate the complex
$R_{II}$ integral.  \citet{tfa71} presented a method of
simplifying the integral using polynomial series expansions.  \citet{fk75}
proposed an even simpler technique, which is still widely used
\citep[e.g.][]{jev81,dgl89}.  However, \citet{gsb80}
questions the accuracy of the Kneer approximation for
supergiant atmospheres, so in our calculations we evaluate $R_{II}$ by
direct numerical integration.

     We adopt a simple Monte Carlo approach for the radiative transfer
calculations.  Individual photons created by collisional excitation are
followed from one scattering event to the next until they escape the
scattering layer.  For each scattering event, we shift to the rest frame of
the scattering layer in order to ensure the frequency redistribution is done
correctly.  For a layer as opaque as the one modeled here,
each photon scatters an average of roughly 10,000 times before escaping.  By
keeping track of where the photons scatter, we gradually build up a
wavelength and height dependent source function, $S_{\lambda}$.  We also keep
track of the trajectory angle $\theta$ of escaping photons relative to the
normal of our plane-parallel atmosphere.  Combining this information and the
source function, we compute line profiles for various values of
$\mu=\cos \theta$ using the familiar radiative transfer equation
\begin{equation}
I_{\lambda}(\tau=0,\mu)=
  \frac{1}{\mu}\int_{0}^{\tau_{\lambda}}S_{\lambda}(t)e^{-t/\mu}dt.
\end{equation}
We then compute a final disk-integrated line profile using a routine from
\citet{jav96}.  The disk integration procedure is really only
valid for thin atmospheres, which will not be the case here.  However, we are
only interested in the gross appearance of the line profile rather than its
exact flux or center-to-limb behavior, so our plane-parallel treatment and
the disk integration procedure just described should be good enough for our
purposes.

\subsection{Results}

     After experimenting with many modest modifications to the assumed
density, temperature, velocity, and microturbulence structure of our model
atmosphere, we settled on the curves shown in Figures 2a-c.
An optically thin Mg~II line produced by the model atmosphere in
Figure 2 has a centroid velocity of about $-15$ km~s$^{-1}$, which is a bit
large but is still within the range of centroid velocities observed for the
Fe~II and Al~II] lines (see \S 1).  The wavelength integrated surface flux
derived for our best-fit model profile is 147 ergs cm$^{-2}$ s$^{-1}$.  If we
convert this into an observed flux assuming a stellar radius of 2 AU and a
stellar distance of 100 pc, which are typical values for the Miras studied in
Paper 1 \citep[see also][]{mk91,mk97,db98}, we find a predicted flux at
Earth of $1.4\times 10^{-12}$ ergs cm$^{-2}$ s$^{-1}$.  This flux is nicely
within the range of observed values at $\phi=0.5$ (see Paper 1).

     Figure 3a shows the model Mg~II h line profile after convolution with
IUE's instrumental profile, which has a width of 0.2 \AA.  Two observed Mg~II
h lines are also shown in Figure 3a for the sake of comparison, one
from R~Car (dotted line) and one from T~Cep (dashed line).  Both are
observations near $\phi=0.5$.  The originally computed model profile was
significantly narrower than the observed profiles, so we had to broaden it
by assuming a macroturbulence of 10 km~s$^{-1}$ in order to obtain the
profile shown in the figure, which is a reasonably good match to the
observations.  The model profile still has a weak red peak, but this is
not grossly inconsistent with observations since the observed R~Car
profile appears to have a similar feature.  The centroid of the model profile
is at about $-33$ km~s$^{-1}$, consistent with the velocities typically
observed around $\phi=0.5$ (see Paper 1).

     It was not an easy matter to find a combination of density,
temperature, velocity, and microturbulence profiles which would suppress the
red peak to the required degree while still maintaining a velocity structure
consistent with the velocities of the optically thin Fe~II and Al~II] lines
and also producing roughly the correct amount of Mg~II flux.  Increasing or
decreasing the assumed densities by a factor of 2, for example, is enough
to significantly increase the flux in the red peak of the model Mg~II line,
degrading the agreement with the observations.  It is not clear at all why
the atmospheric structure of Miras should be so finely tuned, so it is quite
possible that our models are still lacking some characteristic that is
important for determining the observed line profiles.  Nevertheless, the
identification of at least one model that adequately fits the data is enough
to demonstrate that scattering in an atmospheric structure like that shown
in Figure 2, with downflowing material above the shock, can in principle
produce the observed Mg~II profiles without any need for absorption from
overlying material.  We have found we can best reproduce the observed
profiles if the microturbulence of the downflowing material is assumed to be
lower than that of the shocked material responsible for the emission,
although models with $b=6$ km~s$^{-1}$ or $b=1$ km~s$^{-1}$ outside the
shock are not dramatically worse than the best models with
$b=3$ km~s$^{-1}$ outside the shock.

     In Figure 3b, our best-fit profile is compared with two other
model profiles:  a profile computed using CRD rather than PRD
(dotted line), and a profile computed using the ionization equilibrium
densities in Figure 2d (dashed line) instead of our preferred estimates of
non-equilibrium densities.  A more substantial red peak is prominent in each
of these profiles, so they do not agree with the observations as well as
our best model.

     The computations assuming ionization equilibrium do not do as well
simply because there is no Mg~II in the downflowing material above the shock
that can scatter red peak photons (see Fig.\ 2d).  All the magnesium is in
its neutral state due to the low temperatures present there.  It is
interesting to note, however, that a strong blue peak asymmetry results
even in the absence of scattering from downflowing material.  Much of the
suppression of the red peak is therefore due to scattering within the
shock itself.  Most of the emergent photons naturally scatter last in the
near side of the shock, suggesting that the positive velocity gradient at
that location helps suppress the red peak all by itself, although adding
scattering from the downflowing material (i.e., switching to the
non-equilibrium model in Fig.\ 3) does suppress the red peak even further.

     The CRD calculations fail to reproduce the observations because it is
too easy for photons to scatter into the far wings of the line, past the
spectral region where the downflowing material can scatter the photons and
prevent their escape.  Thus, CRD computations do not dramatically suppress
the red peak of the line and the resulting profile is extremely broad.
Previous work has demonstrated that CRD is a very poor approximation for red
giant and supergiant atmospheres and generally produces unrealistically
broad Mg~II lines \citep{gsb80,sad83,dgl89}.

     Although we have succeeded in roughly reproducing the Mg~II line
profiles, there is one possible problem with the model, and that is that it
produces an h line profile that is too narrow.  As mentioned above, we
corrected this by adding macroturbulent line broadening, but the presence of
such macroturbulence may also broaden less opaque lines to a similar extent,
depending on whether the macroturbulence exists where the Mg~II lines are
originally formed or in higher layers of the atmosphere where the emission
is scattered.  In any case, this model may not be able to reproduce the
narrow profiles of the Fe~II and Al~II] lines, which are much less opaque
than Mg~II h \& k if not actually optically thin.  It is possible that the
narrow model profile reflects inaccuracies introduced into the calculation
by the various assumptions that have been made (i.e., the plane-parallel
approximation and crude disk integration technique) rather than the
existence of macroturbulent flow fields.

     Despite the potential difficulty just described, we conclude that it is
possible in principle for atmospheric structures like those predicted by
hydrodynamic models of Miras to account for the lack of red peaks in
observed Mg~II profiles from these pulsating stars.  Further modeling is
needed to confirm these results, and to find a model which accurately
reproduces the profiles of both Mg~II h \& k and of the less optically thick
Fe~II and Al~II] lines.  In addition, future modeling could combine
hydrodynamic modeling and a more sophisticated treatment of the radiative
transfer in a self-consistent manner.

\section{Summary}

     The large blueshifts of the centroids of Mira Mg~II lines are unlikely
to be indicative of mass motions this rapid, because less optically thick
lines suggest more modest flow velocities.  The Mg~II k line is clearly
contaminated by absorption from overlying material, but this contamination
does not explain the large blueshifts, since the blueshifts are seen not
only for the Mg~II h \& k lines but also for the Ca~II H \& K lines, and the
probability of overlying absorption producing similar blueshifts in all four
very opaque lines is very small.  We have used radiative transfer
calculations to determine if shock structures such
as those suggested by published hydrodynamic models of Miras
\citep[e.g.][]{ghb88} can generate Mg~II h line profiles like those observed
by IUE.  In particular, we test whether downflowing material above the
shocks, such as that predicted by the models, can suppress the red side of
the emission lines, thereby producing highly blueshifted profiles like those
that are observed.  We find that scattering in such an atmosphere can
reproduce the observed Mg~II profiles and fluxes under the following
conditions:
\begin{description}
\item[1.] Calculations assuming complete redistribution cannot
  reproduce the observed profiles.  The effects of partial redistribution
  must be taken into account.
\item[2.] Simple estimates of recombination timescales suggest the plasma
  outside the shocks is out of ionization equilibrium, and our attempt to
  correct for these effects improves the agreement between the
  observed and modeled Mg~II profiles.  However, a strong blue peak
  asymmetry results even when equilibrium densities are assumed and
  there is no scattering from the overlying downflowing material.
  This suggests that scattering {\em within} the shock accounts for much of
  the suppression of the red side of the line.
\item[3.] Although the velocity predicted for optically thin lines in our
  model is consistent with observed velocities of Fe~II and Al~II] lines,
  there may be a problem with the line widths.  We had to assume a
  macroturbulence of 10 km~s$^{-1}$ to broaden our model Mg~II profile
  enough to be consistent with observations, but the existence of such
  macroturbulence would possibly broaden the Fe~II and Al~II] lines more
  than is observed.  More sophisticated modeling in the future could
  resolve this issue.
\end{description}

     Understanding the gross properties of the observed Mg~II lines
is a necessary first step before trying to understand the
phase-dependent behavior described in Paper 1.
Phase-dependent changes in density, temperature,
velocity, microturbulence, and macroturbulence may all play a role
in producing this behavior.

\acknowledgments

     We wish to thank the referee, R.\ Robinson, for comments which 
improved the paper.  MK is a member of the Chandra Science Center, which
is operated under contract NAS-839073, and is partially supported by NASA.

\clearpage

\begin{figure}
\plotone{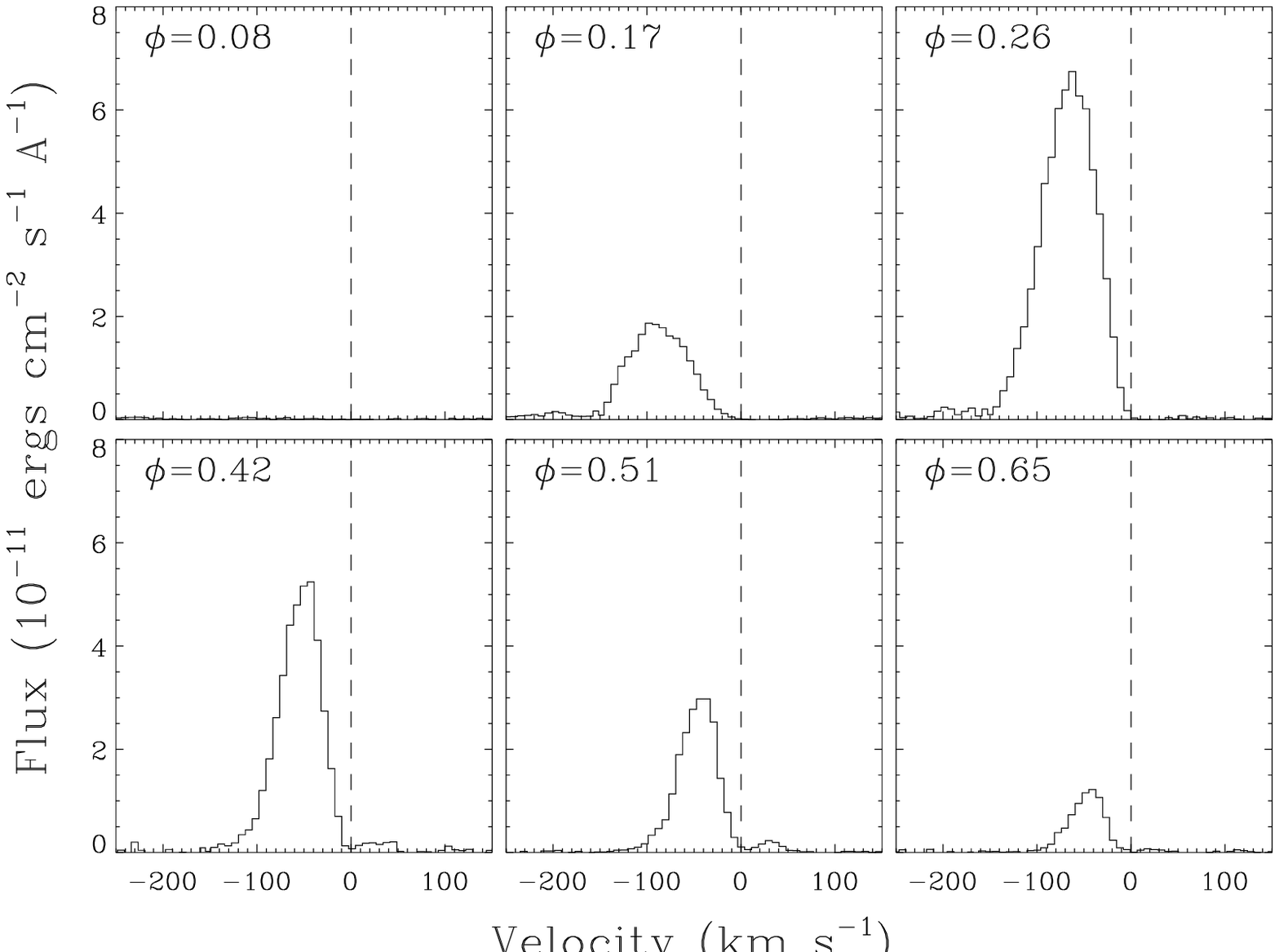}
\caption{A sequence of IUE spectra of the Mg~II h line
  observed in 1989-1990 from the Mira variable R~Car, illustrating how the
  profile typically varies during the course of a pulsation cycle.  The
  spectra are plotted on a velocity scale in the stellar rest frame.}
\end{figure}

\clearpage

\begin{figure}
\plotfiddle{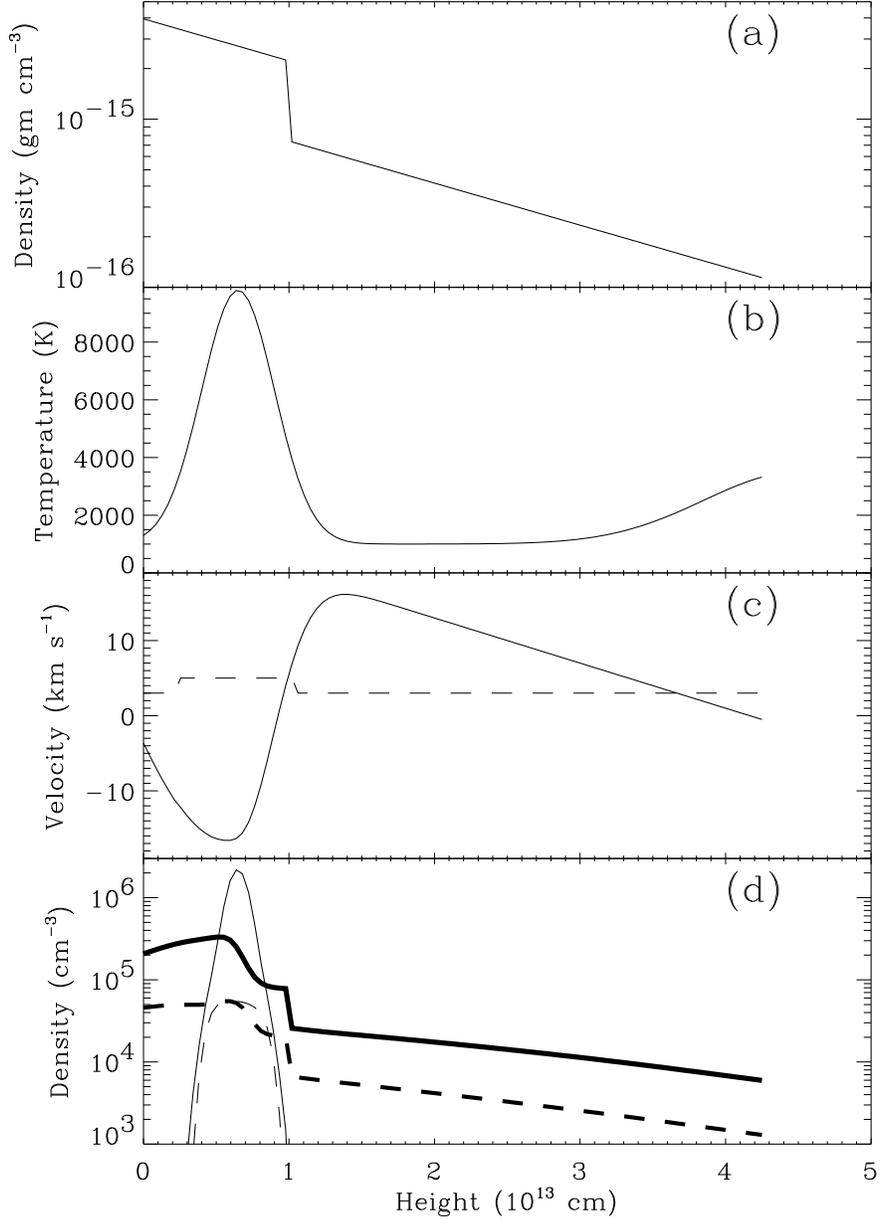}{6.0in}{0}{80}{80}{-320}{-10}
\caption{A schematic model of the innermost shock of a
  pulsating Mira and the region overlying it.  The density, temperature,
  and flow velocity are shown as solid lines in panels (a)-(c).  Panel (c)
  also displays the assumed microturbulent velocity profile for our
  ``best-fit'' radiative transfer model (dashed line).  Panel (d) displays
  two estimates of the electron density (solid lines) and the Mg~II density
  (dashed lines), one which assumes ionization equilibrium (thin lines), and
  one which attempts to correct for non-equilibrium effects (thick lines).}
\end{figure}

\clearpage

\begin{figure}
\plotfiddle{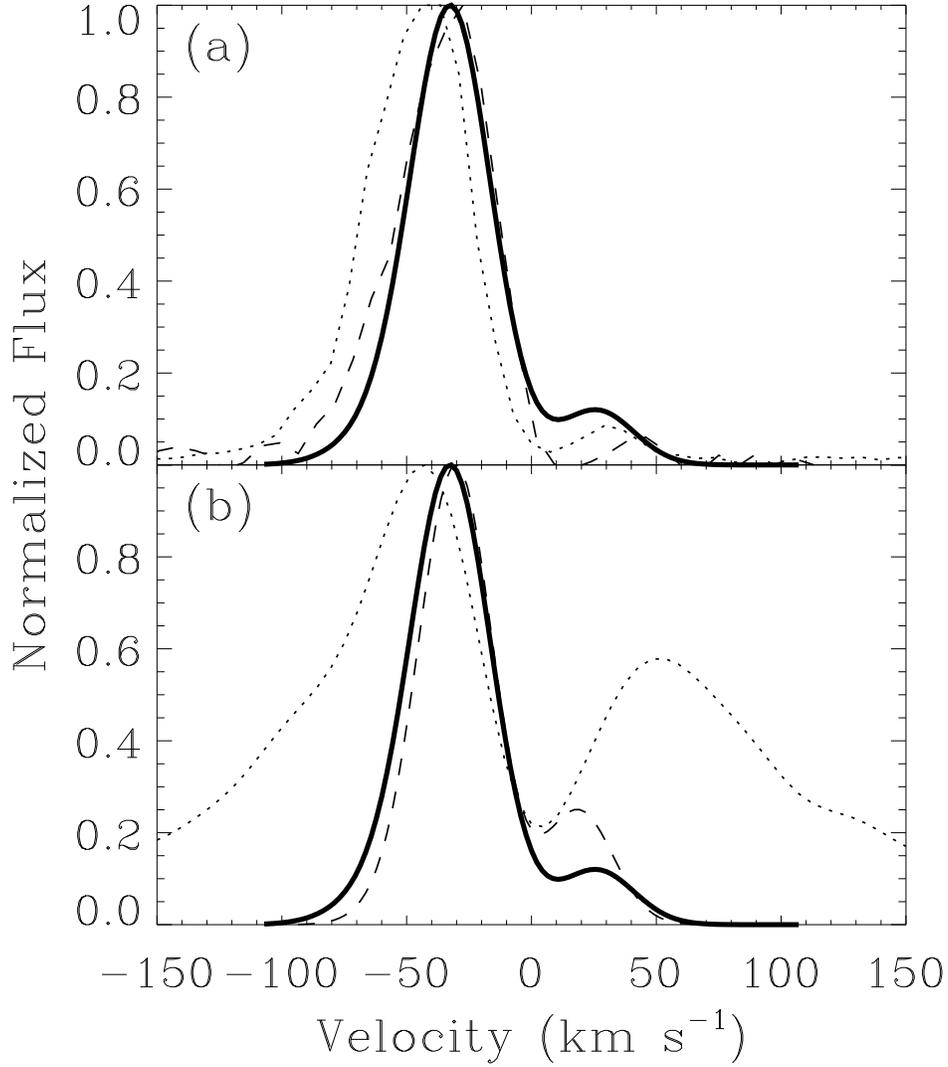}{4.5in}{0}{80}{80}{-270}{0}
\caption{(a) Our ``best-fit'' model Mg~II h line profile (solid line)
  compared with two observed profiles, one from R~Car (dotted line) and one
  from T~Cep (dashed line).  (b) In this panel, the same best-fit profile
  from (a) is compared with two other model profiles:  a profile computed
  using CRD rather than PRD (dotted line), and a profile computed assuming
  ionization equilibrium densities rather than our preferred non-equilibrium
  densities (dashed line).}
\end{figure}

\end{document}